\documentclass[pra,aps]{revtex4}
\usepackage{graphicx}
\usepackage{amsmath}
\numberwithin{equation}{section}

\begin{document}

\title{Algorithmic Complexity in Cosmology and Quantum Gravity }
\author{V. Dzhunushaliev}
\email{dzhun@hotmail.kg}
\affiliation{Dept. Phys.and Microel. Engineer., Kyrgyz-Russian
Slavic University, Bishkek, Kievskaya Str. 44, 720000, Kyrgyz
Republic}

\author{D. Singleton}
\email{dougs@csufresno.edu}
\affiliation{Physics Dept., CSU Fresno, 2345 East San Ramon Ave.
M/S 37 Fresno, CA 93740-8031, USA}

\date{\today}

\begin{abstract}
In this article we use the idea of algorithmic complexity (AC)
to study various cosmological scenarios, and
as a means of quantizing the gravitational interaction.
We look at 5D and 7D cosmological models where the Universe
begins as a higher dimensional Planck size spacetime which
fluctuates between Euclidean and Lorentzian signatures. These
fluctuations are governed by the AC of the two different signatures. 
At some point a transition to a 4D Lorentzian signature
Universe occurs, with the extra dimensions becoming ``frozen'' or
non-dynamical. We also apply the idea of algorithmic complexity
to study composite wormholes, the entropy of blackholes, and the
path integral for quantum gravity.
\end{abstract}

\maketitle

\section{Introduction} 

The modern cosmological paradigm is that Universe started from the
Big Bang, which was the origin not only of all matter and energy,
but also gave rise to the physical laws of Nature:
Einstein gravity, Yang-Mills equations, quantum mechanics {\it etc.}
In this article we examine the possibility that the Big Bang
was a quantum birth ({\it i.e.} a quantum fluctuation) of the Universe
from Nothing. With this view 
one can imagine that there could exist other Universes with
different physical laws ({\it e.g.} non-Einstein gravity). Thus one
would like to assign some probability for a given Universe to fluctuate into
existence. Based on path integral ideas one can write the
probability for a given Universe to come into existence as
\begin{equation}
  P = A\exp{\left(-S\right)}
\label{intr1}
\end{equation}
$S$ is an action which has contributions from the fields that occur
in the given Universe, and the factor $A$ is connected with
the type of physical laws in the Universe. Such an
expression is only valid at or near the Planck scale. 
\par 
These arguments lead to the following assumption: \textit{on the 
Planck scale the physical laws can fluctuate}. This implies
that there is ``something'' that distinguishes one set
of physical laws from another. This ``something''  
influences what kind of Universe with what kind of physical laws 
will appear. Intuitively we expect that the simpler a physical law
({\it e.g.} the field equations)  the more
probable is the corresponding Universe. This
is a free rendering of Einstein's idea that
``Everything should be made as simple as possible, but not simpler.''
The problem is how to recognize or formulate this ``something''.
Our proposal is that this ``something'' is connected
with Kolmogorov's ideas on algorithmic complexity (AC).
In this approach any physical system ({\it e.g.} the Universe)
can be thought of in terms of an algorithm. The longer and
more complex the algorithm, the less likely it is for such
a system to appear. In particular Universes 
with different physical laws (field equations) are described
by different algorithms. The length of these algorithms then
affects the probability that this Universe with a certain set of
physical laws will fluctuate into existence.

The above discussion leads to the idea that the physical laws
of a Universe are in some sense dynamical. We will take the
dynamical nature of the physical laws for different Universes
as non-differentiable or discrete quantity. The
non-differentiable dynamics can have two manifestations :
\begin{itemize}
  \item 
  The cosmological appearance of a Universe with certain
  physical laws.
  \item 
  The quantum fluctuations of physical laws at the level
  of the spacetime foam ({\it e.g.} at the Planck scale).
\end{itemize}
The first case was discussed above -- see Eq.\eqref{intr1}. As an
example of the second case consider a 5D spacetime with a 
mostly non-dynamical $55$ metric component. Thus for most
of spacetime we have 4D gravity + electromagnetism,
\textit{i.e.} 5D Kaluza-Klein theory in its initial
interpretation. However, there are small regions where the
$55$ metric component is
a dynamical variable and one has full 5D gravity. These
fully dynamical regions can be thought of as quantum handles in
the spacetime foam \cite{dzh1}.
 
We will now give mathematical details to this hypothesis
about the connection between algorithmic complexity and the
probability for the existence of a given Universe with certain
fields and certain physical laws.

\section{Kolmogorov's algorithmic complexity}

The mathematical definition for algorithmic complexity (AC) is 
\par
\textit{
The algorithmic complexity $K(x\mid y)$ of the  object $x$  for a 
given object $y$ is the minimal length of the ``program'' $P$
that is written as a sequence of  the  zeros  and  ones
which allows us to construct $x$ starting from $y$:
\begin{equation}
K(x\mid y) = \min_{A( P,y)=x} l(P)
\label{ac1}
\end{equation}
$l(P)$ is length of the  program $P$; $A(P,y)$  is  the
algorithm  for calculating an object $x$, using  the  program $P$,
when the object $y$ is given.} 
\par
This definition gives us an exact mathematical meaning to
the word ``simple'' in the spirit of Einstein's
above-mentioned  statement. In the next few sections we
will demonstrate this idea of the connection
between algorithmic complex and cosmology and gravity
with some examples. 

\section{A toy model for the birth of Minkowski space}

In this section we sketch a model for the emergence of 4D Minkowski 
spacetime from a collapsing 7D spacetime as the result of a quantum
fluctuation. The probability for this transition to occur is linked with
the algorithmic complexity of the equations describing either
the 4D Minkowski spacetime or the empty 7D spacetime \cite{dzh2}.
Since this transition involves a discrete change in the
number of spacetime dimensions it \textit{can not}
be described by classical or quantum field theory. It must
be described by some non-differentiable (discrete)  mechanics.
We start with an empty 7D spacetime with the metric given by
\begin{equation}
ds^{2} = dt^{2} - a^{2}(t) dl^{2}_{1} - b^{2}(t) dl^{2}_{2},
\label{min10}
\end{equation}
where $dl^{2}_1 = dx^2 + dy^2 + dz^2$ is the metric of the
3D flat space $E^{3}$ ; $dl^{2}_2 = du^2 + dv^2 + dw^2$ is the
metric of the extra dimensions (ED) which
are also a flat $E^3$ space. The 7D Lagrangian is \cite{sal}
\begin{equation}
  L = \sqrt{G} R_{7D} = \sqrt{-G} \left( R_{4D} + 
  \left(cross \; \; terms \right) + R_{ED}\right)
\label{min11}
\end{equation}
here $R_{ED} = 0$; $G$ is the determinant of the 7D metric;
$R_{7D}$ and $R_{4D}$ are the scalar curvature of the
7D and 4D spaces respectively. The Einstein
vacuum field equations have the following form 
\begin{subequations}
\begin{eqnarray}
&&\frac {\ddot a}{a} = -\frac {\dot a ^{2}}{a^{2}} +
                     \frac {\dot b ^{2}}{b^{2}},
\label{min20a}\\
&&\frac {\ddot b}{b} = -\frac {\dot b ^{2}}{b^{2}} +
                     \frac {\dot a ^{2}}{a^{2}},
\label{min20b}\\
&&3\frac{\dot a \dot b}{ab} + \frac {\dot a ^{2}}{a^{2}} +
                           \frac {\dot b ^{2}}{b^{2}} = 0,
\label{min20c}
\end{eqnarray}
\label{min20}
\end{subequations}
where  $(\dot{\phantom  x})$ is the derivative with respect
to $t$. This system has the following exact Kazner solution 
\begin{subequations}
\begin{eqnarray}
a & = & a_{0} \left( - \frac{t}{a_{1}} \right) ^{\alpha};
\qquad
b = b_{0} \left( - \frac{t}{b_{1}} \right) ^{\beta};
\qquad t < 0;
\label{min30a}\\
\alpha & = & \frac{1 - \sqrt{5}}{6};
\qquad
\beta = \frac{1 + \sqrt{5}}{6}
\label{min30b}
\end{eqnarray}
\label{min30}
\end{subequations}
where $a_0 \gg b_0 \gg l_{Pl}$ and $a_1 = b_1 = l_{Pl}$
($l_{Pl}$ is the Planck length).
This represents a collapsing 7D spacetime.
The scalar curvature is
\begin{equation}
  -\frac{R}{6} = \frac{\ddot a}{a} + \frac{\ddot b}{b} +
  3\frac{ \dot a \dot b}{ab} +
  \frac{{\dot a}^2}{a^2} + \frac{{\dot b}^2}{b^2}
\label{min40}
\end{equation}
for these constants $a_{0,1}$ and $b_{0,1}$ the Ricci scalar is 
$R \approx 1/l^2_{Pl}$ when $|t| \approx t_{Pl}$. 
\par
At times close to the Planck time ($|t| \approx t_{Pl}$)
we will assume that quantum fluctuations between spacetimes
of different dimensions is more likely. Thus
there should be some likelihood of a
spontaneous transition from a 7D to a 4D spacetime, so that
three of the extra spatial dimensions of the 7D spacetime
become non-dynamical. Mathematically this is written as
\begin{subequations}
\begin{eqnarray}
L_{7D} & \longrightarrow & L_{4D}
\label{min50a}\\
\sqrt{-G} \left( R_{4D} +   \left(cross\;term\right) \right) 
& \longrightarrow & 
\sqrt{-g}R_{4D} ,
\label{min50b}
\end{eqnarray}
\label{min50}
\end{subequations}
where $g$ is determinant of the 4D metric. The form of the
4D metric is a lower dimensional version of the 7D
metric
\begin{equation}
ds^{2} = dt^{2} - a^{2}(t) dl^{2}
\label{min60}
\end{equation}
where the element $dl^{2} = dl^2_1$ from Eq. \eqref{min10}.
Einstein's equations for this metrics are:
\begin{subequations}
\begin{eqnarray}
2\frac{\ddot a}{a} + \frac{\dot a ^{2}}{a^{2}} & = & 0,
\label{min70a}\\
\frac{\dot a^{2}}{a^{2}} & = & 0,
\label{min70b}
\end{eqnarray}
\label{min70}
\end{subequations}
These equations have the following simple solution 
\begin{equation}
a = a'_{0} = const.
\label{min80}
\end{equation}
which is just 4D Minkowski spacetime. The probability for
such a fluctuation to occur is determined by the AC of the
4D versus the 7D case. Looking at equations \eqref{min20} and
\eqref{min70} one can see that the AC of the 7D Universe
is larger than for the 4D Universe since the system of equations
are more complex (the number of equations describing the 7D
case is larger than the number of equations describing the
4D case).
\par
A summary of this idea of the emergence of a lower
dimensional Universe from a higher dimensional one
goes as follows
\begin{itemize}
  \item 
  First, for $t<0$ we have an empty 7D Kazner Universe ($-\infty < t < 0$) 
  evolving according to \eqref{min20}. This solution is collapsing
  toward a singularity at $t = 0$.
  \item 
  Second, at time $|t| \approx t_{Pl}$ a quantum fluctuation of
  the dimensionality of spacetime takes place. This results in
  a quantum splitting
  off of the ED, \textit{i.e.} three of the six spatial dimensions
  from the 7D Universe become
  non-dynamical resulting in an effective 4D Universe.
  \item
  Third, the linear 4D scales ($a_{0}$ for 3D space and $b_{0}$ for
  the other three EDs) become fixed, classical variables whose
  values are determined by the values they took just before the
  splitting off of the EDs. Thus we have a static, 4D, Minkowski
  Universe with three non-dynamical EDs.
\end{itemize}
The probability $P$  for this transition  from
a multidimensional Universe to a $4D$ Universe is
determined by the AC of the two Universes. Mathematically
we can write
\begin{equation}
P_{multiD\to 4D} = \frac{e^{-K_2}}
             {e^{-K_1} + e^{-K_2}},
\label{min90}
\end{equation}
where $K_{1,2}$ are respectively the AC of the
multidimensional and the 4D Universes described by the
algorithms (system of equations) \eqref{min20}
and \eqref{min70}. Since the system \eqref{min20} is larger
({\it i.e.} more complex) than the system \eqref{min70}
we will assume $K_{1}\gg K_{2}$ (even in simple cases the
detailed calculation of AC is a very complicated problem).
Thus Eq.\eqref{min90} can be approximate as follows:
\begin{equation}
P_{multiD\to 4D}\approx 1 - e^{-\left[ K_1 - K_2 
            \right ]} \approx 1.
\label{min100}
\end{equation}

\section{Fluctuation of the metric signature}

In this section we present a variation of Bousso and Hawking's idea 
\cite{hawking} that the Universe began as an Euclidean space, 
\textit{i.e.} spacetime with Euclidean time, and later
evolved into a Universe with Lorentzian signature. The variation
that we want to consider is that the Universe started out
as a multi-dimensional space with its
metric fluctuating between Euclidean and Lorentzian
signatures \cite{dzh4}. At some point on the boundary of this space a
transition to a 4D Universe with a definite signature takes place. 
\par 
Since the metric signature is not a continuous variable its
dynamics can not be described by differential equations.
To see this consider the multi-dimensional metric.
\begin{equation}
ds^2_{(MD)} = \eta_{\bar A\bar B} 
\left (h^{\bar A}_C dx^C \right )
\left (h^{\bar B}_D dx^D \right )
\label{fluct10}
\end{equation}
$\eta_{\bar A\bar B}$ is the signature of the metric with
viel-bein indices $\bar A, \bar B=0,1,2,3,5,6,\cdots$.
$x^A$ are the coordinates on the total space of
the principal bundle with a structural group $\cal G$, and
$C,D$ are the multidimensional (MD) coordinate indices.
The metric on the total space of the principal
bundle (we will consider gravity on the principal bundle) 
can be rewritten
\begin{equation}
ds^2_{(MD)} = \eta_{\bar a\bar b} 
\left ( 
h^{\bar a}_c dx^c + h^{\bar a}_\mu dx^\mu
\right )
\left ( 
h^{\bar b}_c dx^c + h^{\bar b}_\mu dx^\mu
\right ) + 
\eta_{\bar \mu\bar \nu} 
\left ( 
h^{\bar \mu}_\alpha dx^\alpha 
\right )
\left ( 
h^{\bar \nu}_\beta dx^\beta 
\right )
\label{fluct20}
\end{equation}
$\bar a,\bar b$ are the viel-bein indices for the fibre 
of the principal bundle, and $c,d$ are the coordinate indices on 
the fibre; $\bar\mu ,\bar\nu$ and $\alpha,\beta$ play the 
same role for the 4D base of the principal bundle. For the
continuous quantities, $h^{\bar A}_C$, we
have gravitational equations, but $\eta_{\bar{A}\bar{B}}$ are discrete 
(non-differentiable) quantities without dynamical equations. 
Thus the dynamics of
the metric signature, $\eta_{\bar A\bar B}$, can not be described 
by differential equations. We will instead apply a 
quantum-like description for these degrees of freedom. This description
will be stochastic along the general lines of 't~Hooft's proposition 
that quantum gravity may be a stochastic phenomenon
\cite{hooft99}. The gravitational field equations on the principal bundle
are deduced in Appendix \eqref{app1}. In the following subsection \eqref{5D}
we take $\Lambda_{1,2} = 0$.

\subsection{The 5D Fluctuating Universe}
\label{5D}

In this subsection we consider the scenario where at
the origin of the Universe a fluctuation
between Euclidean and Lorentzian metrics occurs. 
This is a modification of an idea initially proposed by
Hawking where there may be regions of the Universe
with Euclidean or Lorentzian signatures. The boundary
between these two regions represents some quantum 
fluctuation between the different metric signatures.
Such transitions between metric signatures could
occur in the very Early Universe on the scale of
Planck length.
\par 
We start with a vacuum 5D Universe with the metric 
\begin{equation}
ds^2_{(5)} = \sigma dt^2 + b(t)\left (d\xi + 
\cos \theta d\varphi \right )^2 + a(r)d\Omega ^2_2 +  
r_0^2 e^{2\psi (t)}\left [d\chi - \omega (t) \left (d\xi + 
\cos \theta d\varphi \right ) \right ]^2 
\label{fluct60}
\end{equation}
here $\sigma = \pm 1$ for the Euclidean and 
Lorentzian signatures respectively. The 3D space metric 
$dl^2 = b(t)\left (d\xi + 
\cos \theta d\varphi \right )^2 + a(r)d\Omega ^2_2$ 
describes the Hopf bundle with an $S^1$ fibre over an 
$S^2$ base. In the 5-bein formalism we have 
\begin{equation}
ds^2_{(5)} = \eta _{\bar A\bar B} e^{\bar A}e^{\bar B}
\label{fluct70}
\end{equation}
here $\bar A, \bar B$ are the 5-bein indices and 
\begin{subequations}
\begin{eqnarray}
\eta _{\bar A\bar B} & = & \left (\pm 1,+1,+1,+1,+1 \right ) ,
\label{fluct80a}\\
e^{\bar 0} & = & dt ,
\label{fluct80b}\\ 
e^{\bar 1} & = & \sqrt b \left (d\xi + 
\cos\theta d\varphi\right ) ,
\label{fluct80c}\\
e^{\bar 2} & = & \sqrt a d\theta ,
\label{fluct80d}\\
e^{\bar 3} & = & \sqrt a \sin\theta d\varphi ,
\label{fluct80e}\\
e^{\bar 5} & = & r_0 e^{\psi}
\left [d\chi - \omega (t) \left (d\xi + 
\cos \theta d\varphi \right ) \right ]
\label{fluct80f}
\end{eqnarray}
\label{fluct80}
\end{subequations}
According to the following theorem \cite{sal} 
\par
\textit{
Let $G$ be a structural group of the principal bundle. Then there is a
one-to-one correspondence between the $G$-invariant metrics
\begin{equation}
ds^2 = \varphi (x^\alpha) \left(\sigma ^a + A^a_\mu dx^\mu \right)^2 + 
g_{\mu\nu}(x^\alpha) dx ^\mu dx^\nu
\label{fluct90}
\end{equation}
on the  total  space ${\cal X}$
and the triples $(g_{\mu \nu }, A^{a}_{\mu }, \varphi )$.
Here $g_{\mu \nu }$ is the 4D metric on the base; $A^{a}_{\mu }$ are the
gauge fields of the group $G$ (the off-diagonal components of the
multidimensional metric); $dl^2 = \sigma^a \sigma_a$ is the symmetric
metric on the fibre $a=5, \cdots , \dim G$ is the index on the fibre and 
$\mu = 0,1,2,3$ is the index on the base.
}
\par
the metric in Eq. \eqref{fluct60} has the following electromagnetic potential
\begin{equation}
A = \omega (t) \left (d\xi + 
\cos \theta d\varphi \right ) = 
\frac{\omega}{\sqrt b} e^{\bar 1}
\label{fluct100}
\end{equation}
For this potential the Maxwell tensor is
\begin{equation}
F = dA = \frac{\dot \omega}{\sqrt b} e^{\bar 0} \wedge e^{\bar 1} - 
\frac{\omega}{a}e^{\bar 2} \wedge e^{\bar 3}
\label{fluct110}
\end{equation}
which yields an electrical field like
\begin{equation}
E_{\bar 1} = F_{\bar 0 \bar 1} =  \frac{\dot \omega}{\sqrt b}
\label{fluct120}
\end{equation}
and a magnetic field like
\begin{equation}
H_{\bar 1} = \frac{1}{2}\epsilon_{1\bar j\bar k}F^{\bar j \bar k} = 
- \frac{\omega}{a}
\label{fluct130}
\end{equation}
The 5D, vacuum Einstein equations \cite{dzh4} resulting from
Eq. \eqref{fluct60} are
\begin{subequations}
\begin{eqnarray}
G_{\bar 0\bar 0} \propto 
2\frac{\dot b \dot \psi}{b} + 4 \frac{\dot a \dot \psi}{a} + 
2\frac{\dot a \dot b}{ab} + \frac{\dot a^2}{a^2} + 
\sigma \left (\frac{b}{a^2} - \frac{4}{a}\right ) + 
r_0^2e^{2\psi}\left (\sigma H_{\bar 1}^2 - E_{\bar 1}^2 
\right ) & = & 0 ,
\label{fluct140a}\\
G_{\bar 1\bar 1} \propto 4\ddot\psi + 4{\dot\psi}^2 + 
4\frac{\ddot a}{a} + 4\frac{\dot a\dot\psi}{a} + 
\sigma\left (3\frac{b}{a^2} - \frac{4}{a} \right ) - 
\frac{{\dot a}^2}{a^2} + 
r_0^2e^{2\psi}\left (\sigma H_{\bar 1}^2 - E_{\bar 1}^2 
\right ) & = & 0 ,
\label{fluct140b}\\
G_{\bar 2\bar 2} = G_{\bar 3\bar 3} \propto 
4\ddot\psi + 4{\dot\psi}^2 + 
2\frac{\ddot b}{b} + 2\frac{\dot b\dot\psi}{b} - 
\frac{\dot b^2}{b^2} + 2\frac{\ddot a}{a} + 
2\frac{\dot a\dot\psi}{a} + 
\frac{\dot a\dot b}{ab} - \frac{\dot a^2}{a^2} - 
\sigma\frac{b}{a^2} - 
r_0^2e^{2\psi}\left (\sigma H_{\bar 1}^2 - E_{\bar 1}^2 
\right ) & = & 0 ,
\label{fluct140c}\\
R_{\bar 5\bar 5} \propto \ddot \psi + {\dot\psi}^2 + 
\frac{\dot a\dot\psi}{a} + \frac{\dot b\dot\psi}{2b} + 
\frac{r_0^2}{2}e^{2\psi}\left (
\sigma H_{\bar 1}^2 + E_{\bar 1}^2
\right ) &= & 0 ,
\label{fluct140d}\\
R_{\bar 2\bar 5} \propto \ddot \omega + \dot\omega
\left (\frac{\dot a}{a} - \frac{\dot b}{2b} + 3\dot\psi
 \right )
-\sigma \frac{b}{a^2}\omega & = & 0
\label{fluct140e}
\end{eqnarray}
\label{fluct140}
\end{subequations}
where $G_{\bar A\bar B} = R_{\bar A\bar B} - \frac{1}{2}\eta_{\bar A\bar B}R$ 
is the Einstein tensor. Our basic assumption is that 
\textit{at the Planck scale there can exist regions where
a quantum fluctuation between Euclidean and Lorentzian metric signatures
occurs}. There are two copies of the classical equations \eqref{fluct140}:
one with $\sigma = +1$ and another with $\sigma = -1$. It is this
quantity $\sigma$ which we take as having quantum fluctuations between
its two discrete values. The basic question under this assumption
is how to calculate the relative probability for each pair of equations
from \eqref{fluct140} (the ones with $\sigma = +1$ versus the ones with
$\sigma = -1$).
\par
We will define the probability for each pair of equations in terms
of the algorithmic complexity of each pair. We can diagrammatically
represent the fluctuations between the Euclidean and Lorentzian
versions of Einstein's equations in the following way
\begin{equation}
\begin{array}{ccc}
\sigma = +1 & \longleftrightarrow & \sigma = -1
\\
& \Downarrow  & 
\\
\left(G^+\right)_{\bar 0\bar 0} 
& 
\longleftrightarrow
& 
\left(G^-\right)_{\bar 0\bar 0} 
\\
\left(G^+\right)_{\bar 1\bar 1} 
& 
\longleftrightarrow
& 
\left(G^-\right)_{\bar 1\bar 1}
\\
\left(G^+\right)_{\bar 2\bar 2} 
& 
\longleftrightarrow
& 
\left(G^-\right)_{\bar 2\bar 2}
\\
\left(G^+\right)_{\bar 3\bar 3} 
& 
\longleftrightarrow
& 
\left(G^-\right)_{\bar 3\bar 3}
\\
\left(R^+\right)_{\bar 5\bar 5} 
& 
\longleftrightarrow
& 
\left(R^-\right)_{\bar 5\bar 5}
\end{array}
\label{fluct150}
\end{equation}
The signs $\pm$ indicates if the equation belongs to the
Euclidean or Lorentzian mode.
Expression \eqref{fluct150} sums up the idea that treating
$\sigma$ as a quantum quantity leads to quantum fluctuations between
the classical equations:
$\left(R^+\right)_{\bar A\bar B} \leftrightarrow
\left(R^-\right)_{\bar A\bar B}$ or $\left(G^+\right)_{\bar A\bar B}
\leftrightarrow \left(G^-\right)_{\bar A\bar B}$.
The probability connected with each pair of equations
($R^\pm_{\bar A\bar B}$  or $G^\pm_{\bar A\bar B}$) 
\textit{is determined by the AC of each equation.} 

\paragraph{\textbf{\underline{Fluctuation $\left(R^+\right)_{\bar 2\bar 5} 
\longleftrightarrow \left(R^-\right)_{\bar 2\bar 5}$.}}} 

The $R_{\bar 2\bar 5}$ equation in the Euclidean and Lorentzian modes is 
respectively 
\begin{subequations}
\begin{eqnarray}
\ddot \omega + \dot \omega\left ( \frac{\dot a}{a} - 
\frac{\dot b}{2b} + 3\dot \psi \right )
-\frac{b}{a^2} \omega = 0 ,
\label{fluct160a} \\
\ddot \omega + \dot \omega\left ( \frac{\dot a}{a} - 
\frac{\dot b}{2b} + 3\dot \psi \right )
+ \frac{b}{a^2} \omega = 0 .
\label{fluct160b}
\end{eqnarray}
\label{fluct160}
\end{subequations}
Let us consider the $\psi = 0$ case (below we 
will see that this is consistent with the
$R_{\bar 5\bar 5}$ equation). It is easy to see that Eq.
\eqref{fluct160a} can be deduced from the instanton condition 
\begin{equation}
E^2_{\bar 1} = H^2_{\bar 1} \qquad or \qquad 
\frac{\omega}{a} = \pm \frac{\dot\omega}{\sqrt b}
\label{fluct180}
\end{equation}
The second equation \eqref{fluct160b} does not have
a similar simplification via the instanton condition
\eqref{fluct180}. This is just the well known fact that instantons
can exist only in Euclidean space. Based of this simplification
from a second order equation \eqref{fluct160a} to a first order equation
\eqref{fluct180} we consider the Euclidean equation \eqref{fluct160a}
simpler from an algorithmic point of view than the Lorentzian 
equation \eqref{fluct160b}. To a first, rough approximation 
we can take the probability of the Euclidean 
mode as $p^+_{25} \approx 1$ and for the Lorentzian mode as $p^-_{25}
\approx 0$. Strictly the exact definition for
each $p^\pm_{ab}$ is  
\begin{equation}
p^\pm_{ab} = \frac{e^{-K^\pm_{ab}}}
{e^{-K^+_{ab}} + e^{-K^-_{ab}}}
\label{fluct190}
\end{equation}
where $K^\pm_{ab}$ is the AC for the $R^\pm_{ab} = 0$ or $G^\pm_{ab} = 0$ 
equation. For $K^+_{25} \ll K^-_{25}$ we have 
$p^+_{25} \approx 1$ and $p^-_{25} \approx 0$. 

The expression for the probability in Eq. \eqref{fluct190} can be
seen as the discrete variable analog of the Euclidean path
integral transition probability. For a continuous variable the
Euclidean path integral gives the probability for the variable
to evolve from some initial configuration to some
final configuration as being proportional to the
exponential of minus the action ($\propto e^{-S}$).
Eq. \eqref{fluct190} is similar, but with the AC replacing
the action. The denominator normalizes the probability
(it is a sum rather than integral since we are dealing
with a discrete variable).

\paragraph{\textbf{\underline{Fluctuation $\left(R^+\right)_{\bar 5\bar 5} 
\longleftrightarrow \left(R^-\right)_{\bar 5\bar 5}$.}}} 

The $R_{\bar 5\bar 5}$ equation in the Euclidean and Lorentzian modes is 
respectively 
\begin{subequations}
\begin{eqnarray}
\ddot \psi + {\dot\psi}^2 + 
\frac{\dot a}{a}\dot\psi + \frac{\dot b}{b}\dot\psi + 
\frac{r_0^2}{2}e^{2\psi}\left (
H^2_{\bar 1} + E^2_{\bar 1}
\right ) =  0 ,
\label{fluct200a}\\
\ddot \psi + {\dot\psi}^2 + 
\frac{\dot a}{a}\dot\psi + \frac{\dot b}{b}\dot\psi + 
\frac{r_0^2}{2}e^{2\psi}\left (
-H^2_{\bar 1} + E^2_{\bar 1}
\right ) =  0 ,
\label{fluct200b}
\end{eqnarray}
\label{fluct200}
\end{subequations}
The Lorentzian mode \eqref{fluct200b} has a trivial solution
\begin{equation}
\psi = 0
\label{fluct220}
\end{equation}
provided the instanton condition ({\it i.e.} $H^2_{\bar 1}
= E^2_{\bar 1}$) holds. Thus for this equation we take
the Lorentzian mode as having a smaller AC, and
in the contrast with the previous subsection,
the Lorentzian mode has the greater probability. Again
to a first, rough approximation the
probability of the Euclidean mode is $p^+_{55} \approx 0$ and 
consequently for the Lorentzian mode $p^-_{55} \approx 1$ . 

\paragraph{\textbf{\underline{Fluctuation $\left(G^+\right)_{\bar 1\bar 1} 
\longleftrightarrow \left(G^-\right)_{\bar 1\bar 1}$ 
and $G^+_{\bar 2\bar 2} \longleftrightarrow G^-_{\bar 2\bar 2}$}}}

Taking into account \eqref{fluct220} we can write these
equations as 
\begin{subequations}
\begin{eqnarray}
4\frac{\ddot a}{a} + 
\sigma\left (3\frac{b}{a^2} - \frac{4}{a} \right ) - 
\frac{{\dot a}^2}{a^2} + 
r_0^2 \left (\sigma H_{\bar 1}^2 - E_{\bar 1}^2 
\right ) & = & 0 ,
\label{fluct240a}\\
2\frac{\ddot b}{b}  - 
\frac{\dot b^2}{b^2} + 2\frac{\ddot a}{a} + 
\frac{\dot a\dot b}{ab} - \frac{\dot a^2}{a^2} - 
\sigma\frac{b}{a^2} - r_0^2
\left (\sigma H_{\bar 1}^2 - E_{\bar 1}^2
\right ) & = & 0 .
\label{fluct240b}
\end{eqnarray}
\label{fluct240}
\end{subequations}
For the Euclidean mode ($\sigma = +1$) with the instanton condition 
\eqref{fluct180}) one can have $b = a$ (an isotropic Universe) which
reduces the two equations of \eqref{fluct240} to \textit{only} one equation
\begin{equation}
4\frac{\ddot a}{a} - \frac{{\dot a}^2}{a^2} - \frac{1}{a} = 0 .
\label{fluct250}
\end{equation}
For the Lorentzian mode ($\sigma = -1$) $b \ne a$ 
(an anisotropic Universe) there are still two equations 
\begin{subequations}
\begin{eqnarray}
4\frac{\ddot a}{a} - 
\left (3\frac{b}{a^2} - \frac{4}{a} \right ) - 
\frac{{\dot a}^2}{a^2} - 
r_0^2 \left (H_{\bar 1}^2 + E_{\bar 1}^2 
\right ) & = & 0 ,
\label{fluct260a}\\
2\frac{\ddot b}{b}  - 
\frac{\dot b^2}{b^2} + 2\frac{\ddot a}{a} + 
\frac{\dot a\dot b}{ab} - \frac{\dot a^2}{a^2} + 
\frac{b}{a^2} + 
r_0^2 \left (H_{\bar 1}^2 + E_{\bar 1}^2 
\right ) & = & 0 ,
\label{fluct260b}
\end{eqnarray}
\label{fluct260}
\end{subequations}
Thus under the instanton condition \eqref{fluct180} and $\psi =0$ we
find that the Euclidean mode \eqref{fluct250} effectively
reduces to one, second order equation which corresponds to an
isotropic Universe; the Lorentzian mode \eqref{fluct260} still has
two, second order equations which describe an
anisotropic Universe. Thus we assign the Euclidean mode the smaller
AC and as for the previous equations make the rough approximation
$p^+_{11} \approx 1$ for the Euclidean mode,
$p^-_{11} \approx 0$ for the Lorentzian mode.

\paragraph{\textbf{\underline{Fluctuation $\left(G^+\right)_{\bar 0\bar 0} 
\longleftrightarrow \left(G^-\right)_{\bar 0\bar 0}$}}} 

The equation  $G^\pm_{\bar 0\bar 0} = 0$ has the
following form
\begin{equation}
2\frac{\dot b \dot \psi}{b} + 4 \frac{\dot a \dot \psi}{a} + 
2\frac{\dot a \dot b}{ab} + \frac{\dot a^2}{a^2} + 
\sigma \left (- \frac{4}{a} + \frac{b}{a^2} \right ) + 
r_0^2e^{2\psi}\left (\sigma H_{\bar 1}^2 - E_{\bar 1}^2 
\right ) = 0
\label{fluct270}
\end{equation}
Assuming all the previous conditions (the instanton condition,
$\psi = 0$, and $b = a$) the Euclidean mode equations become
\begin{equation}
\frac{\dot a^2}{a^2} - \frac{1}{a} = 0 
\label{fluct280} 
\end{equation}
while the Lorentzian mode equations become
\begin{equation}
3\frac{\dot a^2}{a^2} + 3 \frac{1}{a} -
r_0^2 \left (H_{\bar 1}^2 + E_{\bar 1}^2 
\right )= 0 .
\label{fluct290}
\end{equation}
The instanton condition again implies that the Euclidean
mode has a smaller AC. Thus to a first, rough 
approximation we take $p^+_{00} \approx 1$ 
and $p^-_{00} \approx 0$. 

\paragraph{\textbf{\underline{Mixed system of the equations}}} 

Under the approximation where the probability associated
with each of the equations in \eqref{fluct140} is
$p \approx 0$ or $1$ the \textit{mixed} system of equations
which describe a Universe \textit{fluctuating between Euclidean
and Lorentzian modes}
\begin{subequations}
\begin{eqnarray}
\frac{\dot a^2}{a^2} - \frac{1}{a} & = & 0 , 
\label{fluct300a}\\
\dot \omega & = & \pm \frac{\omega}{\sqrt a} ,
\label{fluct300b}\\
4\frac{\ddot a}{a} - \frac{{\dot a}^2}{a^2} 
- \frac{1}{a} & = & 0 .
\label{fluct300c}
\end{eqnarray}
\label{fluct300}
\end{subequations}
here $b = a$, $\psi = 0$ and the instanton condition are all
assumed to hold. This system of mixed Euclidean and Lorentzian
equations has the following simple solution
\begin{subequations}
\begin{eqnarray}
a & = & \frac{t^2}{4} ,
\label{fluct310a}\\
\omega & = & t^2 .
\label{fluct310b}
\end{eqnarray}
\label{fluct310}
\end{subequations}

\paragraph{\textbf{\underline{The mixed origin of the Universe}}}
\label{mixed}

The following model for the quantum birth of 
Universe has been advanced by Hawking :
one begins with an Euclidean space of the Planck size
($R^4$, $S^4$ or some other smooth non-singular Euclidean space); 
then a Lorentzian Universe emerges from a boundary of this 
initial Euclidean piece. In this scenario the  
Euclidean and Lorentzian spaces are connected by a hypersurface
with a mixed signature.
\par
In this section we present a variation of this picture for the
quantum mechanical origin of the Universe. We assume that the
Universe begins as a quantum fluctuating system between Euclidean and
Lorentzian modes. Then at some point in time there is
a quantum transition to the Lorentzian mode. 
\par
To support these statements mathematically we begin by
calculating the average of the Ricci scalar
\begin{equation}
  \biggl\langle R(\sigma )\biggl\rangle = p^+ R\left(\sigma = +1\right) + 
  p^- R\left(\sigma = -1\right)
\label{fluct311}
\end{equation}
where $p^+$ and $p^-$ are the probabilities for the scalar curvature with 
$\sigma = +1$ and $\sigma = -1$ respectively. Using 
\begin{equation}
  -\frac{3}{2} R(\sigma) = G^{\bar{\alpha}}_{\bar{\alpha}} + R^{\bar{5}}_{\bar{5}}
\label{fluct312}
\end{equation}
and averaging gives
\begin{eqnarray}
  -\frac{3}{2} \biggl\langle R(\sigma) \biggl\rangle & = & p^+_{\alpha \alpha} 
  \left( G^+ \right)^{\bar{\alpha}}_{\bar{\alpha}} + 
  p^-_{\alpha\alpha} 
  \left( G^- \right)^{\bar{\alpha}}_{\bar{\alpha}} + 
  p^+_{55} \left( R^+ \right)^{\bar{5}}_{\bar{5}} + 
  p^-_{55} \left( R^- \right)^{\bar{5}}_{\bar{5}} \nonumber \\
  & = &
  \left( G^+ \right)^{\bar 0}_{\bar 0} + 
  \left( G^+ \right)^{\bar 1}_{\bar 1} + 
  \left( G^+ \right)^{\bar 2}_{\bar 2} + 
  \left( G^+ \right)^{\bar 3}_{\bar 3} + 
  \left( R^- \right)^{\bar 5}_{\bar 5} .
\label{fluct313}
\end{eqnarray}
Thus for the mixed system of equations we find
\begin{equation}
  \biggl\langle R (\sigma) \biggl\rangle = 0 .
\label{fluct317}
\end{equation}
In this toy model the Universe originates from an
empty, multidimensional,  non-singular (in the sense that 
$\langle R (\sigma)\rangle = 0$), spacetime of Planck
scale size ($\tau \lesssim \tau_{Pl}$). In our model
the spacetime is $M^4 \times S^1$, with $M^4$ being a
space with fluctuating metric signature:
Euclidean~$\leftrightarrow$~Lorentzian. At some point
a quantum transition to the Lorentzian mode occurs, and
at the same or later time the $55$ component of the metric
becomes a non-dynamical quantity. Thus the fluctuation
of the metric signature of the original Planck scaled, 5D
Universe leads to a 4D Lorentzian Universe {\it{and}} a
``frozen'' or non-dynamical 5$^{th}$ dimension. 

\subsection{The 7D Fluctuating Universe}

In this subsection we study a 7D cosmological solution with a
fluctuating metric signature as in the last subsection. We take 
the gauge group of the EDs as ${\cal G} = SU(2)$, with the
7D metric taking the form
\begin{equation}
ds^2 = b\left( x^\alpha \right)
\left (\omega^{\bar a} + A^{\bar a}_\mu \left( x^\alpha \right)dx^\mu\right )
\left (\omega_{\bar a} + A_{\bar a\mu} \left( x^\alpha \right)dx^\mu\right ) + 
g_{\mu\nu}\left( x^\alpha \right) dx^\mu dx^\nu .
\label{b0}
\end{equation}
Most of the calculational details for this 7D metric are given in 
Appendix \eqref{app1}.
\par 
The total space of the principal bundle is denoted as $E$; the
structural group is denoted as $\cal G$. 
The factor-space ${\cal H}=E/{\cal G}$ is the base of the principal
bundle, and is described by the 4D metric
\begin{equation}
ds^2_{(4)} = \eta_{\bar \mu\bar \nu} 
\left ( 
h^{\bar \mu}_\alpha dx^\alpha 
\right )
\left ( 
h^{\bar \nu}_\beta dx^\beta 
\right )
\label{b1}
\end{equation}
which is the last term in Eq. \eqref{b0}. We now insert a 4D cosmological
constant term into the MD action
\begin{equation}
S = \int \left (R + 2\Lambda_1 \right )
\sqrt{|G|}d^{4+N}x + 
\int 
\left (2\Lambda_2\right )
\sqrt{|g|}d^4x 
= \int 
\left [ 
\int \left (R + 2\Lambda_1 \right )
\sqrt{|\gamma|}d^Ny + 2\Lambda_2
\right ]
\sqrt{|g|}d^4x
\label{b2}
\end{equation}
$R$ is the Ricci scalar and $G_{AB} = \eta_{\bar C\bar D}h^{\bar C}_Ah^{\bar D}_B$ 
is the MD metric on the total space; 
$g_{\mu\nu} = \eta_{\bar\alpha\bar\beta}h^{\bar\alpha}_\mu
h^{\bar\beta}_\nu$ is the 4D metric on the base of the principal 
bundle;  $\gamma_{ab} = \eta_{\bar c\bar d}h^{\bar c}_ah^{\bar d}_b$ 
is the metric on $\cal G$; $G, g$ and 
$\gamma$ are the appropriate metric determinates; 
$\Lambda_{1,2}$ are the MD and 4D $\Lambda$-constants; 
$N=dim({\cal G})$. The MD action of Eq. \eqref{b2} has several
points in common with the 4D EYM action considered in
Ref. \cite{hoso} (non-zero cosmological constants and
effective SU(2) ``Yang-Mills'' gauge fields).  Eq. \eqref{b2}
also has a connection to the action for the Non-gravitating Vacuum
Energy Theory \cite {gund}. In Ref. \cite{gund} Guendelman considers
an action which has degrees of freedom which are independent of the
metric, with the resulting action having two measures of integration
(involving metric and non-metric degrees of freedom). Eq. \ref{b2}
incorporates two distinct degrees of freedom : the continuous variables,
$h^{\bar A}_B$, and the discrete variables, $\eta_{\bar A\bar B}$.
In Ref. \cite{gund} both the metric and non-metric degrees of
freedom were continuous.
\par 
The independent, \textit{continuous degrees} of freedom  
are: the vier-bein $h^{\bar\mu}_\nu(x^\alpha)$, 
the gauge potential $h^{\bar a}_\mu(x^\alpha) = A^{\bar a}_\mu(x^\alpha)$ 
and the scalar field $b(x^\alpha)$. $e^{\bar a}_b$ is defined as 
\begin{equation}
\omega^{\bar a} = e^{\bar a}_b dx^b
\label{b6}
\end{equation}
$x^b$ are the coordinates on the group $\cal G$; 
$\omega^{\bar a}$ are the 1-forms satisfying
\begin{equation}
d\omega^{\bar a} = f^{\bar a}_{\bar b\bar c}
\omega^{\bar b}\wedge\omega^{\bar c}
\label{b7}
\end{equation}
$f^{\bar a}_{\bar b\bar c}$ are the structural constants 
of SU(2). Varying the action in Eq. \eqref{b2} with respect to
$h^{\bar\mu}_\nu$, $h^{\bar a}_\nu$ and
$b$ leads to (see the Appendix for details) 
\begin{subequations}
\begin{eqnarray}
R_{\bar\mu\bar\nu} - \frac{1}{2}\eta_{\bar\mu\bar\nu} R & = & 
\eta_{\bar\mu\bar\nu}
\left (\Lambda_1 + \frac{\Lambda_2}{b^{3/2}} \right ) ,
\label{b8}\\
R_{\bar a\bar\mu} & = & 0 ,
\label{b9}\\
R^{\bar a}_{\bar a} & = & - \frac{6}{5}
\left (\Lambda_1 + \frac{\Lambda_2}{b^{3/2}} \right ) .
\label{b10}
\end{eqnarray}
\label{b10a}
\end{subequations}
Eq. \eqref{b8} are the Einstein vacuum equations with $\Lambda$-terms;
Eq. \eqref{b9} are the ``Yang-Mills'' equations; Eq. \eqref{b10}is
reminiscent of Brans-Dicke theory since the metric on each fibre is
symmetric and has only one degree of freedom - the scalar factor $b(x^\mu)$. 
\par
We now investigate Eqs. \eqref{b8}-\eqref{b10} using the ansatz
\begin{equation}
ds^2 = \sigma dt^2 + b(t)
\left (\omega^{\bar a} + A^{\bar a}_\mu dx^\mu\right )
\left (\omega_{\bar a} + A_{\bar a\mu} dx^\mu\right ) + 
a(t)d\Omega^2_3
\label{b11}
\end{equation}
$\sigma=\pm1$ describes the possible quantum fluctuation 
of the metric signature between Euclidean and Lorentzian modes, 
$A^{\bar a}_\mu$ are SU(2) gauge potentials, 
$d\Omega^2_3=d\chi^2 + \sin^2\chi\left (
d\theta^2 + \sin^2\theta d\phi^2 \right )$ is the metric on the 
unit $S^3$ sphere and $x^0=t, x^1=\chi ,x^2=\theta ,
x^3=\phi , x^5=\alpha ,x^6=\beta ,x^7=\gamma$. 
($\alpha ,\beta ,\gamma$ are the Euler angles for the SU(2) 
group) 
\begin{subequations}
\begin{eqnarray}
\omega^1 & = & \frac{1}{2}
(\sin \alpha d\beta - \sin \beta \cos \alpha d\gamma ),
\label{b12}\\
\omega^2 & = & -\frac{1}{2}(\cos \alpha d\beta +
\sin \beta \sin \alpha d\gamma ),
\label{b13}\\
\omega^3 & = & \frac{1}{2}(d\alpha +\cos \beta d\gamma ).
\label{b14}
\end{eqnarray}
\label{b14a}
\end{subequations}
The off-diagonal components of the MD metric 
take the instanton-like form \cite{instanton} \cite{don} 
\begin{subequations}
\begin{eqnarray}
A^a_\chi & = & \frac{1}{4}\left \{ -\sin\theta \cos\varphi ;
-\sin\theta \sin\varphi ;\cos\theta \right \}
(f(t) - 1),
\label{b16}\\
A^a_\theta & = & \frac{1}{4}\left \{ -\sin\varphi ;
-\cos\varphi ;0\right \}(f(t) - 1),
\label{b17}\\
A^a_\varphi & = & \frac{1}{4}\left \{0;0;1\right \}
(f(t) - 1).
\label{b18}
\end{eqnarray}
\label{b18a}
\end{subequations}
Substituting into Eqs. \eqref{b8}-\eqref{b10} gives 
\begin{subequations}
\begin{eqnarray}
\frac{1}{3}R^{\bar a}_{\bar a} = R^{\bar 5}_{\bar 5} = 
-\frac{\sigma}{2}\frac{\ddot b}{b} + \frac{2}{b} - 
\frac{\sigma}{4}\frac{{\dot b}^2}{b^2} - 
\frac{3}{4}\sigma \frac{\dot a\dot b}{ab} + 
\frac{1}{8}\frac{b}{a}
\left (\sigma E^2 + H^2 \right )
& = & -\frac{2}{5}\left (\Lambda_1 + 
\frac{2\Lambda_2}{b^{3/2}} \right ) ,
\label{b19}\\
G_{\bar 0\bar 0} = 
-3\frac{\sigma}{b} + \frac{3}{4}\frac{{\dot b}^2}{b^2} - 
3\frac{\sigma}{a} + 
\frac{9}{4}\frac{\dot a\dot b}{ab} + 
\frac{3}{16}\frac{{\dot a}^2}{a^2} - 
\frac{3}{16}\frac{b}{a}\left (E^2 - \sigma H^2 \right ) 
& = & \sigma \left (\Lambda_1 + 
\frac{\Lambda_2}{b^{3/2}} \right ) ,
\label{b20}\\
G_{\bar 1\bar 1} = \frac{3}{2}\sigma\frac{\ddot b}{b} - 
\frac{3}{b} + \sigma \frac{\ddot a}{a} - 
\frac{1}{a} + \frac{3}{2}\sigma\frac{\dot a\dot b}{ab} - 
\frac{\sigma}{4}\frac{{\dot a}^2}{a^2} + 
\frac{1}{16}\frac{b}{a}\left (\sigma E^2 - H^2 \right ) 
& = & \left (\Lambda_1 + 
\frac{\Lambda_2}{b^{3/2}} \right ) ,
\label{b21}\\
G_{\bar 2\bar 7} = 
2\ddot f + 5\frac{\dot b\dot f}{b} + 
\frac{\dot a\dot f}{a} - 
4\frac{\sigma}{a}f\left (f^2 - 1 \right ) 
& = & 0 ,
\label{b22}\\
E^2 = E^a_i E^{ai} = \dot f^2 , 
\quad 
H^2 = H^a_i H^{ai} & = & \frac{
\left (f^2 - 1 \right )^2}{a} ,
\label{b23}
\end{eqnarray}
\label{b23a}
\end{subequations}
$G_{\bar A\bar B} = R_{\bar A\bar B} - 
(1/2)\eta_{\bar A\bar B} R$; $i=1,2,3$ are space indices;
the ``electromagnetic'' fields are
\begin{equation}
E^a_i = F^a_{0i}, \quad H^a_i = 
\frac{1}{2}\varepsilon_{ijk} F^{ajk}
\label{b24}
\end{equation}
$F^a_{\mu\nu}$ is the field strength tensor for the non-Abelian
gauge group. The wormhole instanton of Ref. \cite{hoso}
had a vanishing ``electric'' field. In contrast the
solution studied here has both non-vanishing ``electric''
and ``magnetic'' fields.

As in the 5D case we assume a quantum fluctuation
between Euclidean and Lorentzian modes which can be
described by a diagram similar to Eq. \eqref{fluct150})
\begin{equation}
\begin{array}{ccc}
\sigma = +1 & \longleftrightarrow & \sigma = -1
\\
& \Downarrow  & 
\\
\left (R^+\right )^{\bar 5}_{\bar 5} 
& 
\longleftrightarrow
& 
\left (R^-\right )^{\bar 5}_{\bar 5} 
\\
\left (G^+\right )_{\bar 0\bar 0} 
& 
\longleftrightarrow
& 
\left (G^-\right )_{\bar 0\bar 0}
\\
\left (G^+\right )_{\bar 1\bar 1} 
& 
\longleftrightarrow
& 
\left (G^-\right )_{\bar 1\bar 1}
\\
\left (G^+\right )_{\bar 2\bar 7} 
& 
\longleftrightarrow
& 
\left (G^-\right )_{\bar 2\bar 7}
\end{array}
\label{qf1}
\end{equation}
As in the 5D case we will estimate the probability for
each pair of equations in \eqref{qf1}.

\paragraph{\textbf{\underline{Fluctuation $\left(G^+\right)_{\bar 2\bar 7} 
 \longleftrightarrow \left(G^-\right)_{\bar 2\bar 7}$}}}

This equation in the Euclidean mode is 
\begin{equation}
2\ddot f + 5\frac{\dot b\dot f}{b} + 
\frac{\dot a\dot f}{a} - 
\frac{4}{a}f\left (f^2 - 1 \right )  = 0 
\label{in1}
\end{equation}
which has the instanton solution 
\begin{equation}
\dot f = \frac{1 - f^2}{\sqrt a} ,
\label{in2}
\end{equation}
where
\begin{equation}
b = b_0 = const
\label{in3}
\end{equation}
Eq. \eqref{in2} implies the instanton condition 
\begin{equation}
E^a_i E_a^i = H^a_i H_a^i .
\label{in4}
\end{equation}
In the Lorentzian mode
\begin{equation}
2\ddot f + 5\frac{\dot b\dot f}{b} + 
\frac{\dot a\dot f}{a} + 
\frac{4}{a}f\left (f^2 - 1 \right )  = 0 
\label{in6}
\end{equation}
and the instanton solution \eqref{in4} is not a solution of \eqref{in6},
since the non-singular, instanton solution exists only in the Euclidean case.
Thus in terms of the AC criteria the Euclidean equation \eqref{in1} 
is simpler than Lorentzian equation \eqref{in6}, since it 
is equivalent to the first order differential equation \eqref{in2}. 
\par
To a first, rough approximation we set
the probability of the $G_{\bar 2\bar 7}=0$ equation for
the Euclidean mode to $p^+_{27} \approx 1$ and the Lorentzian mode to 
$p^-_{27} \approx 0$. The exact definition for each $p^\pm_{AB}$ probability
is given in Eq. \eqref{fluct190}. If $K^+_{27} \ll K^-_{27}$ we have 
$p^+_{27} \approx 1$ and $p^-_{27} \approx 0$.

\paragraph{\textbf{\underline{Fluctuation $\left(R^+\right)^{\bar 5}_{\bar 5} 
 \longleftrightarrow \left(R^-\right)^{\bar 5}_{\bar 5}$}}}

This equation in the Euclidean and Lorentzian modes is respectively 
\begin{subequations}
\begin{eqnarray}
-\frac{1}{2}\frac{\ddot b}{b} + \frac{2}{b} - 
\frac{1}{4}\frac{{\dot b}^2}{b^2} - 
\frac{3}{4}\frac{\dot a\dot b}{ab} + 
\frac{1}{8}\frac{b}{a}
\left (E^2 + H^2 \right ) = 
-\frac{2}{5}\left (\Lambda_1 + 
\frac{2\Lambda_2}{b^{3/2}} \right ) ,
\label{a1}\\
\frac{1}{2}\frac{\ddot b}{b} + \frac{2}{b} + 
\frac{1}{4}\frac{{\dot b}^2}{b^2} + 
\frac{3}{4}\frac{\dot a\dot b}{ab} + 
\frac{1}{8}\frac{b}{a}
\left (-E^2 + H^2 \right ) = 
-\frac{2}{5}\left (\Lambda_1 + 
\frac{2\Lambda_2}{b^{3/2}} \right ) ,
\label{a2}
\end{eqnarray}
\label{a12}
\end{subequations}
The Lorentzian mode equation is simpler because the two last terms 
annihilate as a consequence of the instanton condition \eqref{in4}. 
To a first rough approximation we set the probability of the
$R^{\bar 5}_{\bar 5}$ equation for the Euclidean mode to $p^+_{55}
\approx 0$ and the Lorentzian mode to $p^-_{55} \approx 1$. 

\paragraph{\textbf{\underline{Fluctuation $\left(G^+\right)_{\bar 0\bar 0} 
 \longleftrightarrow \left(G^-\right)_{\bar 0\bar 0}$}}}

This equation in the Euclidean mode and Lorentzian mode is respectively 
\begin{subequations}
\begin{eqnarray}
-\frac{3}{b} + \frac{3}{4}\frac{{\dot b}^2}{b^2} - 
\frac{3}{a} + 
\frac{9}{4}\frac{\dot a\dot b}{ab} + 
\frac{3}{16}\frac{{\dot a}^2}{a^2} - 
\frac{3}{16}\frac{b}{a}\left (E^2 - H^2 \right ) = 
\left (\Lambda_1 + \frac{\Lambda_2}{b^{3/2}} \right ) 
\label{tt1}\\
\frac{3}{b} + \frac{3}{4}\frac{{\dot b}^2}{b^2} + 
\frac{3}{a} + 
\frac{9}{4}\frac{\dot a\dot b}{ab} + 
\frac{3}{16}\frac{{\dot a}^2}{a^2} - 
\frac{3}{16}\frac{b}{a}\left (E^2 + H^2\right ) = 
-\left (\Lambda_1 + \frac{\Lambda_2}{b^{3/2}} \right ) .
\label{tt2}
\end{eqnarray}
\label{tt12}
\end{subequations}
In this case because of the instanton condition \eqref{in4} the Euclidean
equation is simpler and therefore in the first rough approximation we can 
set the probability of the $G_{\bar 0\bar 0}=0$ equation for 
the Euclidean mode to $p^+_{00} \approx 1$ and the 
Lorentzian mode to $p^-_{00} \approx 0$. 

\paragraph{\textbf{\underline{Fluctuation $\left(G^+\right)_{\bar 1\bar 1} 
 \longleftrightarrow \left(G^-\right)_{\bar 1\bar 1}$}}}

This equation in the Euclidean mode and Lorentzian mode is respectively 
\begin{subequations}
\begin{eqnarray}
\frac{3}{2}\frac{\ddot b}{b} - 
\frac{3}{b} + \frac{\ddot a}{a} - 
\frac{1}{a} + \frac{3}{2}\frac{\dot a\dot b}{ab} - 
\frac{1}{4}\frac{{\dot a}^2}{a^2} + 
\frac{1}{16}\frac{b}{a}\left (E^2 - H^2 \right ) = 
\left (\Lambda_1 + \frac{\Lambda_2}{b^{3/2}} \right ) 
\label{cc1}\\
-\frac{3}{2}\frac{\ddot b}{b} - 
\frac{3}{b} - \frac{\ddot a}{a} - 
\frac{1}{a} - \frac{3}{2}\frac{\dot a\dot b}{ab} + 
\frac{1}{4}\frac{{\dot a}^2}{a^2} - 
\frac{1}{16}\frac{b}{a}\left (E^2 + H^2 \right ) = 
\left (\Lambda_1 + \frac{\Lambda_2}{b^{3/2}} \right ) .
\label{cc2}
\end{eqnarray}
\label{cc12}
\end{subequations}
As in the previous paragraph, as a consequence of the instanton condition
\eqref{in4}, the Euclidean mode is simpler. Therefore in the first rough
approximation we set $p^+_{11} \approx 1$ and $p^-_{11} \approx 0$. 

\paragraph{\textbf{\underline{Mixed system of equations}}}

The mixed system of equations for the 7D spacetime 
with fluctuating metric signature is 
\begin{subequations}
\begin{eqnarray}
2\ddot f + 5\frac{\dot b\dot f}{b} + 
\frac{\dot a\dot f}{a} - 
\frac{4}{a}f\left (f^2 - 1 \right ) & = & 0 ,
\label{m1}\\
\frac{1}{2}\frac{\ddot b}{b} + \frac{2}{b} + 
\frac{1}{4}\frac{{\dot b}^2}{b^2} + 
\frac{3}{4}\frac{\dot a\dot b}{ab} + 
\frac{1}{8}\frac{b}{a}
\left (-E^2 + H^2 \right ) & = & 
-\frac{2}{5}\left (\Lambda_1 + 
\frac{2\Lambda_2}{b^{3/2}} \right ) ,
\label{m2}\\
-\frac{3}{b} + \frac{3}{4}\frac{{\dot b}^2}{b^2} - 
\frac{3}{a} + 
\frac{9}{4}\frac{\dot a\dot b}{ab} + 
\frac{3}{16}\frac{{\dot a}^2}{a^2} - 
\frac{3}{16}\frac{b}{a}\left (E^2 - H^2 \right ) & = & 
\left (\Lambda_1 + \frac{\Lambda_2}{b^{3/2}} \right ) ,
\label{m3}\\
\frac{3}{2}\frac{\ddot b}{b} - 
\frac{3}{b} + \frac{\ddot a}{a} - 
\frac{1}{a} + \frac{3}{2}\frac{\dot a\dot b}{ab} - 
\frac{1}{4}\frac{{\dot a}^2}{a^2} + 
\frac{1}{16}\frac{b}{a}\left (E^2 - H^2 \right ) & = & 
\left (\Lambda_1 + \frac{\Lambda_2}{b^{3/2}} \right )  .
\label{m4}
\end{eqnarray}
\label{m4a}
\end{subequations}
The solution for this system is
\begin{subequations}
\begin{eqnarray}
a & = & t^2 ,
\label{m5}\\
f & = & \frac{t^2 - t_0^2}{t^2 + t_0^2} ,
\label{m6}\\
b & = & b_0 = const ,
\label{m7}\\
\Lambda_1 & = & -\frac{1}{b_0} ,
\label{m8}\\
\Lambda_2 & = & -2\sqrt b_0 .
\label{m9}
\end{eqnarray}
\label{m9a}
\end{subequations}
The existence of this solution is somewhat surprising ! 
Normally in any dimension the Bianchi identities are satisfied.
Therefore some gravitational field equations are not independent of
the others. Ordinarily the superfluous equations are associated with
initial conditions (\textit{i.e.} Eq. \eqref{m3} above). In our case
the mixed system above comes from a model with a varying metric signature.
As a consequence the  Bianchi identities are not correct and 
this system should be unsolvable. Evidently the 
solution is a condition for the solvability of the mixed system 
which uniquely defines the $\Lambda$-constants. 
If the solution in Eqs. \eqref{m9a} is unique then it
must be absolutely stable. 
\par 
The physical meaning of this solution is: 
\begin{itemize}
\item
Eq. \eqref{m5} implies a flat 4D Einstein spacetime 
that is not effected by matter.
\item
Eq. \eqref{m6} implies a Polyakov - 't Hooft instanton
gauge field configuration which is not effected by gravity.
\item 
Eq. \eqref{m7} implies a frozen ED. 
\item
Eqs. \eqref{m8}-\eqref{m9} imply that the dynamical 
equations uniquely determine the $\Lambda_{1,2}$-constants.
\end{itemize}
It is interesting to note that the effective cosmological
constant terms on the right hand side of Eqs. \eqref{b8}
\eqref{b10} ({\it i.e.} $\Lambda_1$ and $\Lambda_2 /b^{3/2}$)
are inversely proportional to the size of the ED, $b_0$. Thus
in order to have a small cosmological constant term one needs
to have a large ED. This could be seen as supporting 
the large extra dimensions scenarios \cite{hamed}.

\subsection{Physical applications of the solutions}

\subsubsection{Regular Universe}

We can interpret the 5D and 7D solutions as a 4D Universe with fluctuating
metric signature, filled with a U(1) and SU(2) instanton gauge field
and frozen ED. Surprisingly this Universe has only one manifestation of
gravity: the frozen ED that result from the fluctuating metric 
signature. These model Universes are simple examples of possible
effects connected with the dynamics of non-differentiable variables.

\subsubsection{Non-singular birth of the Universe}

Various researchers ({\it e.g.} see Ref. \cite{hawking})
have speculated about the quantum birth
of the Universe from ``Nothing''. In light of this we can
interpret a small piece (with linear size of the
Planck length $\approx l_{Pl}$) of
our model 5D/7D Universe as a quantum birth of the regular 
4D Universe. In contrast to other scenarios 
this origin has a metric signature trembling between 
Euclidean and Lorentzian modes. Further we postulate that
on a boundary of this spacetime there occurs 
\begin{itemize}
\item
\textit{a quantum transition} to only one Lorentzian mode 
with a fixed metric signature.
\item
\textit{a splitting off} the ED so that the metric on the 
fibres $(h^{\bar a}_b)$ becomes a non-dynamical variable. 
After this splitting off the linear size of the gauge group remains 
constant yielding ordinary 4D Einstein-Yang-Mills gravity. 
\end{itemize}
These assumptions about a quantum transition from
fluctuating metric signature $(\pm 1,+1, \cdots ,+1)$ 
to Lorentzian signature $(-1,+1, \cdots ,+1)$ and
a splitting off of the ED should not be seen as something
extraordinary and new, but rather as an extension of our
postulate about the quantum birth of the regular 4D
Universe, discussed above, with certain laws
(gravitational equations + non-differentiable
dynamic). The present case can be seen a quantum-stochastic  
change or evolution of these laws (here this involves only the quantum
transition of $\eta_{00}$ and the splitting off of the ED). 
\par
The probability for the quantum birth is 
\begin{equation}
P \approx N e^{-S}
\label{ns1}
\end{equation}
where $S$ is the Euclidean, dimensionless action, which should be
$S \approx 1$ in Planck units. The factor $N$ is of more interest,
since it contains information about the topological
structure of the boundary of the origin. 
\par 
The probability for the quantum-stochastic transition to Lorentzian 
mode and splitting off of the ED should be determined by the AC of
the final and initial states. Such a quantum-stochastic transition 
can occur only if the final state with Lorentzian mode and splitting 
off of the ED is simpler than the initial state with the fluctuating 
metric signature and dynamic ED.

\section{Algorithmic complexity applied to non-cosmological systems}

In the following three subsections we give examples of the application
of algorithmic complexity to various non-cosmological systems.
First, we study a composite wormhole which consists of a 5D
throat region connecting two Reissner-Nordstr\"om blackholes.
Second, we estimate the entropy of the simplest vacuum solution
to 4D gravity: the Schwarzschild black hole. Finally, we look at the
path integral in gravity in terms of AC.

\subsection{A composite 5D wormhole as the sum of Holographic principle 
and the AC idea}

In this section we construct a composite wormhole 
by connecting two 4D Reissner-Nordstr\"om solutions via
a 5D wormhole-like throat. There are two holographic surfaces 
located between the two Reissner-Nordstr\"om and 5D solution. For the 
Reissner-Nordstr\"om solution the surface is an event horizons,
and for the 5D solution the surface is a $T-$horizon
(the properties of $T-$horizons is discussed below).
The main idea of this subsection is that such a composite object is simpler
in terms of AC than either component separately.
This follows from the fact that the Reissner-Nordstr\"om solution has
a very complicated time dependent metric under the event horizon
whereas the 5D throat does not. In contrast, outside the event
horizon the 4D Reissner-Nordstr\"om solutions is simpler than
the 5D throat.

We begin by considering the 5D wormhole-like metric 
\begin{equation}
ds^{2} = {\Delta (r)}dt^{2} - 
dr^{2} - a(r)d\Omega ^2 + 
- r_1^2 \Delta (r)(d\chi - \omega (r)dt)^2 
\label{fluct30}
\end{equation}
$\chi $ is the extra, 5$^{th}$ coordinate. The metric is symmetric
around $r = 0$. The 5D vacuum Einstein equations are
\begin{subequations}
\begin{eqnarray}
\frac{\Delta ''}{\Delta} - \frac{{\Delta '}^2}{\Delta ^2} +  
\frac{a' \Delta '}{a\Delta} + r_1^2 \Delta ^2{\omega '}^2 & = & 0 ,
\label{fluct40a}\\
\omega '' + 2 \omega ' \frac{\Delta '}{\Delta} + 
\omega ' \frac{a'}{a} & = & 0 ,
\label{fluct40b}\\
\frac{{\Delta '}^2}{\Delta ^2} + \frac{4}{a} - 
\frac{{a'}^2}{a^2} - r_1^2 \Delta ^2{\omega '}^2 & = & 0 ,
\label{fluct40c}\\
a'' - 2& = & 0 
\label{fluct40d}
\end{eqnarray}
\label{fluct40}
\end{subequations}
These equations have the following solution \cite{chodos} \cite{dzhsin}  
\begin{subequations}
\begin{eqnarray}
a & = & r^{2}_{0} + r^{2},
\label{fluct50a}\\
\Delta & = & \frac{q}{2r_0}\frac{r^2 - r_0^2}{r^2 + r_0^2} ,
\label{fluct50b}\\
\omega & = &  \frac{4r_0^2}{r_1q}\frac{r}
{r^2 - r_0^2} 
\label{fluct50ac}
\end{eqnarray}
\label{fluct50}
\end{subequations}
Where $r_0 > 0$ and $q$ are constants.
\par
The composite wormhole that we consider \cite{dzh3}
consists of two 4D Reissner-Nordstr\"om black holes
which are connected by the wormhole-like solution of
\eqref{fluct50}. One interpretation for this composite
wormhole is as a model of a quantum handle in the spacetime foam.
The 5D and 4D physical quantities must be ``sewn'' together by
the following conditions:
\begin{subequations}
\begin{eqnarray}
\frac{1}{\Delta_0} - r_1^2 \omega ^2_0 \Delta_0 = G_{tt}\left 
(\pm r_0\right ) = g_{tt}\left (r_+\right ) = 0,
\label{fluct51a}\\
a_0 = G_{\theta\theta}(\pm r_0) = g_{\theta\theta}(r_+) = r^2_+,
\label{fluct51b}
\end{eqnarray}
\label{fluct51}
\end{subequations}
$G$ and $g$ are 5D and 4D metric tensors respectively, and 
$r_+$ is the event horizon for the Reissner - Nordstr\"om 
solution. The quantities with the $(0)$ subscript are
evaluated at $r=\pm r_0$. Note that $G_{tt}(\pm r_0) = 0$ 
and $ds^2 = 0$ on the surfaces $r = \pm r_0$. 
Hypersurface such as $r = \pm r_0$ have been
called $T-$horizons by Bronnikov \cite{bron}.
\par
$G_{\chi t}$ can be connected to the 4D electric field by examining
the 5D ($R_{\chi t}=0$) and 4D Maxwell equations 
\begin{subequations}
\begin{eqnarray}
\left [a^2\left (\omega '\Delta^2 \right )\right ]' = 0,
\label{fluct52a}\\
\left (r^2E_r\right )' = 0,
\label{fluct52b}
\end{eqnarray}
\label{fluct52}
\end{subequations}
$E_r$ is the 4D electric field.
These two equations are essentially Gauss's law; they indicate that
some quantity multiplied by an area is conserved. In 4D this quantity 
is the 4D Maxwell electric field. We can naturally 
join this 4D, Reissner - Nordstr\"om electric
field, $E_{RN} = e/r^2_+$, with the Kaluza - Klein, ``electrical"  field, 
$E_{KK} = \omega '\Delta^2$, on the event and $T-$horizons 
\begin{equation}
\omega _0'\Delta^2_0 = \frac{q}{a_0} = E_{KK} = 
E_{RN} = {\frac{e}{r^2_+}}.
\label{fluct53}
\end{equation}
It is interesting to note that the event and $T-$horizons can be viewed as 
holographic surfaces which can be used to define the whole spacetime 
\cite{dzh5} \cite{soloduk}. To show this we
consider the 4D and 5D metrics in turn. The metric for the
Reissner-Nordstr\"om spacetime is
\begin{equation} 
ds^2 = \delta (r) dt^2 - \frac{dr^2}{\delta (r)} - 
r^2 \left (d\theta ^2 + \sin ^2 \theta d\varphi ^2 
\right ) , 
\label{2-1-1} 
\end{equation} 
and the electromagnetic potential is 
\begin{equation} 
A_\mu = \{\omega (r), 0, 0, 0\} .
\label{2-1-2} 
\end{equation} 
The Einstein - Maxwell equations are 
\begin{eqnarray} 
-\frac{\delta '}{r} + \frac{1 - \delta}{r^2} & = &
\frac{\kappa}{2} {\omega '} ^2  ,
\label{2-1-3} \\
-\frac{\delta ''}{2} - \frac{\delta '}{r} & = & 
-\frac{\kappa}{2} {\omega '} ^2  ,
\label{2-1-4} \\
\omega ' & = & \frac{q}{r^2}     .
\label{2-1-5}
\end{eqnarray} 
Eq. \eqref{2-1-4} is a consequence of \eqref{2-1-3}  and \eqref{2-1-5}.
For the Reissner - Nordstr\"om blackhole the event horizon is
defined by the condition $\delta (r_g) = 0$, where $r_g$ is the radius of
the event horizon. Hence in this case we see that on the event horizon 
\begin{equation} 
\delta '_g = \frac{1}{r_g} - \frac{\kappa}{2}
r_g {\omega '_g}^2  ,
\label{2-1-6} 
\end{equation}
here (g) means that the corresponding value is evaluated
on the event horizon. Thus the Einstein equation, Eq. \eqref{2-1-3},
is a first-order differential equation 
in the spacetime outside the horizon $(r \ge r_g)$.
Condition \eqref{2-1-6} tells us that the derivative of the
metric on the event horizon is expressed
through the value of the metric on the event horizon. This
shows that the Holographic principle applies in this case
since the spacetime can be determined from information on some
surface (the event horizon).
\par 
Now we consider the 5D WH-like metric \eqref{fluct30} with field equations 
\eqref{fluct40}. On the $T-$horizon $\Delta (\pm r_0) = 0$, and
therefore from Eq. \eqref{2-1-3} we have
\begin{equation}
\Delta '_0 = \pm \frac{q}{a_0} = \pm \frac{q}{2r_0^2}
\label{3-1-8} .
\end{equation}
The signs $(\pm)$ correspond, respectively, to $(r = \mp r_0)$ where the 
$T-$horizons are located. This also indicates that the Holographic
principle applies to the $T-$horizons. 
\par
From an algorithmic point of view we can now argue that such
a composite structure is more likely to occur than either
the 4D Reissner-Nordstr\"om or 5D wormhole solution separately.
The AC of the interior, throat region of the composite system is
calculated with one algorithm (the 5D Einstein vacuum equations) while
the AC of the exterior region is calculated with another
algorithm (the 4D Einstein-Maxwell equations). Because the AC of the
interior region can be calculated from the Holographic principle using
the $T$-horizon, its AC is simpler than if it had been
calculated from the 4D Einstein-Maxwell equations algorithm
especially since the metric under the event horizon is time dependent.
The AC of the exterior region can be calculated from the
Holographic principle using the event horizon, with the 4D
Einstein-Maxwell equations as the algorithm. Thus the exterior
regions AC is simpler if it is calculated using the 4D Einstein-Maxwell
equations rather than the 5D vacuum Einstein equations.
Applying the Holographic principle and ideas of AC we have found
that the composite wormhole has a lower AC than either solution
separately. Such a composite wormhole is only expected to
be important at the Planck scale.

\subsection{The AC of the Schwarzschild black hole.}
\label{algor}

Beckenstein \cite{bek1} and Hawking \cite{haw} have shown 
that an entropy can be associated with a black hole. The
entropy is connected the area of the black hole's event horizon.
Usually the concept of entropy arises in statistical  systems
where one has a great number of particles. However, in the
case of the entropy of a black hole one
associates an entropy with a single object ({\it i.e.} the
black hole). In some sense AC is a concept similar to entropy.
In this section we will estimate the AC for
a Schwarzschild black hole.
\par
For some gravitational field configuration the AC is
determined, according to the definition  \eqref{ac1},
by the smallest  algorithm which yields this configuration
({\it i.e.} which yields the metric). Thus the Einstein
equations are the algorithm for calculating the gravitational
field configuration. In order to calculate the metric for the
whole spacetime one must have, in addition to Einstein's
field equations, some initial  and/or  boundary  conditions.
\par
In order to estimate the AC for the  Schwarzschild  black
hole we write the metric in the following form:
\begin{equation}
ds^{2} = dt^{2} - e^{\lambda (t,R)}dR^{2} - r^{2}(t,R) 
\left( d\theta ^{2} + \sin ^{2} \theta d\phi ^{2} \right),
\label{41}
\end{equation}
for which the Einstein's equations are:
\begin{subequations}
\begin{eqnarray}
-e^{-\lambda}{r'}^2 + 2r\ddot r + \dot r ^2 + 1 & = & 0,
\label{42a}\\
-\frac{e^{-\lambda}}{r}\left (2r'' - r'\lambda ' \right ) + 
\frac{\dot r \dot \lambda}{t} + \ddot \lambda + 
\frac{\dot \lambda ^2}{2} + 
\frac{2\ddot r}{r} & = & 0,
\label{42b}\\
-\frac{e{-\lambda}}{r^{2}}
 \left(2rr'' + {r'} ^{2} -rr'\lambda '\right) +
 \frac{1}{r^{2}}\left( r\dot a \dot \lambda + {\dot a} ^{2} + 
1 \right) & = & 0,
\label{42c}\\
2\dot r' - \dot \lambda r' & = & 0,
\label{43d}
\end{eqnarray}
\end{subequations}
where $(')$ and $(\dot{\phantom x})$ are respectively derivatives
of $t$ and $r$. We take the $t=0$ section as a Cauchy  hypersurface.  
The initial data on this hypersurface then defines the metric 
on the whole  Schwarzschild spacetime. However, because of the
Holography principle the amount of initial
data needed is smaller than one would naively expect.
From  Eq.\eqref{42c} one sees that for $t=0$ the first time
derivative of all components of the metric tensor are zero.
Therefore the initial data must satisfy:
\begin{equation}
2rr''  + {r'} ^{2} - rr' \lambda '  - e^{\lambda } = 0.
\label{43}
\end{equation}
In order to solve Eq. \eqref{43} on the surface $t=0$  we take
boundary conditions of the following form:
\begin{equation}
r' (R=0,t=0) = 0,\qquad r(R=0,t=0) = r_{g},
\label{44}
\end{equation}
where $r_{g}$ is radius at the event horizon. Thus the metric
on the whole Schwarzschild spacetime is defined by the value
of the $G_{\theta\theta}$ component of metric at the origin.
The AC for the Schwarzschild metrics can be written as the sum of 
two quantities. The first quantity is connected with some 
Lorentz-invariant number which is related to the event horizon 
(the surface $t=0, R=0$). The second quantity is connected with 
the Einstein equations. We take the first quantity to
be related to the area of the event horizon ($4\pi r_g^2$).
We will divide this by $4\pi l_{Pl}^2$ in order to
obtain a dimensionless number. The second quantity is taken as
the length of the program for calculating the metric. Thus the AC 
of the Schwarzschild black hole is given by the following expression :
\begin{equation}
K \approx  L\left[ \left (\frac{r_{g}}{l_{Pl}} \right) ^2 \right ]
+ L_{Einstein}, 
\label{46}
\end{equation}
$L[(r_g/l_{Pl})^2]$ is  the program length for the  
definition  of the dimensionless number $r^{2}_{g}/l^{2}_{Pl}$
which is determined from some  universal machine. $L_{Einstein}$ is the 
program length  of the solution of Einstein's differential equations using 
some universal machine, for example, the Turing machine. Finding
an exact expression for the length, $L$, for determining the number
$(r_g/l_{Pl})^2$ is a difficult problem. As a rough approximation
we assume that each Planck sized cell, $l_{Pl} ^3$ can contain
one bit so that $L[(r_g/l_{pl})^2] \approx (r_g/l_{pl})^2$.
With this approximation we can compare the first term of
Eq. \eqref{46} with the Beckenstein-Hawking equation
\begin{equation}
  S = 4\pi r_g^2.
\label{47}
\end{equation}
Thus there appears to be some relation between these two
quantities.

\subsection{Algorithmic complexity and the path integral}

In this section we will propose an alternative method of
calculating the path integral in quantum gravity. The basic idea
is to replace the action ($I[g]$) in the path integral by
the AC ($K[g]$). It is important to note that $K[g]$ is a positive
functional of $g$. Under this replacement of the action by the AC
the path integral becomes
\begin{equation}
\int  D[g] e^{-i(I[g] + \int g_{\mu\nu}J^{\mu\nu}dx)} \rightarrow  
\int  D[g] e^{-i(K[g] + \int g_{\mu\nu}J^{\mu\nu}dx)} = 
e^{iZ[J^{\mu\nu}]},
\label{21a}
\end{equation}
where $g_{\mu\nu}$ is some arbitrary metric; $K[g]$ is the AC for
the metric $g$; $Z[J]$ is a generating functional for quantum gravity.
\par
The most complicated gravitational fields (in terms of AC)
are those metrics which satisfy or are the result of no
field equations. Such configurations are essentially
random fields with no algorithm connecting the values of the
metric at neighboring points in the spacetime. Thus
according to Kolmogorov's definition of AC such random metrics
would have a large AC. Metrics which are the solutions
to some gravity  equations  (Einstein's equations, $R^{2}$  -
theory, Euclidean theory, {\it etc.}) have a smaller 
AC in comparison with random metrics. In this sense one
can take gravitational instantons as the simplest gravitational
objects: they are symmetrical spaces, with
the corresponding metrics possessing the same symmetry group.
One way of understanding why instantons have a small AC is that
they can be determined via their topological charges rather than
by the field equations. This greatly reduces the AC of
such configurations.
\par
Thus, as a first approximation the path integral in quantum gravity
can be defined as the sum over the gravitational instantons. The
next order of approximation would include the contributions from metrics
which are solutions of Einstein's  equations, $R^{2}$  -  theories,
multidimensional  theories {\it etc.} The larger the AC of a
given configuration the larger the order of approximation at
which it contributes to the path integral. An interesting point is
that for quantum gravity based on the integral \eqref{21a} the
Universe can contain different regions where different gravitational
equations hold. An example of this is the composite wormhole discussed
above.

\section{Conclusions}

In this paper we have considered the possibility that Nature
can have changing the physical laws. We have postulated
that the dynamics of this changing may be connected with the AC
of a particular set of laws. This leads to the proposition that
\textit{an object with a  smaller AC has a greater probability to
fluctuate into existence.}
\par 
Some physical consequences that can results from this hypothesized
fluctuation of physical laws at the Planck scale are:
the birth of the Universe with a fluctuating metric signature;
the transition from a fluctuating metric signature to Lorentzian one;
``frozen'' extra dimensions as a consequence of this transition;
quantum handles in the spacetime foam as regions with
multidimensional gravity and so on.

\appendix
\section{Gravitational equations}
\label{app1}

We start from the Lagrangian adopted for the vacuum gravitational 
theory on the principal bundle with the structural group 
$\cal G$ ($\dim ({\cal G}) = N$). $\cal G$ is the gauge group
associated with the EDs
\begin{equation}
S = \int \left (R + 2\Lambda_1 \right )
\sqrt{|G|}d^{4+N}x + 
\int 
\left (2\Lambda_2'\right )
\sqrt{|g|}d^4x 
\label{ap1}
\end{equation}
where $R$ is the Ricci scalar for the total space; $G$ 
and $g$ are the determinant of the metric on the total 
space and base of the principal bundle respectively, 
$\Lambda_1, \Lambda_2'$ are the MD and 4D $\lambda$-constants. This 
Lagrangian is correct if the coordinate transformations 
conserve the topological structure of the total space 
({\it i.e.} does not mix the fibres)
\begin{subequations}
\begin{eqnarray}
{y'}^a & = & {y'}^a \left (y^b\right ) + 
f^a\left ( x^\alpha \right ) ,
\label{ap2}\\
{x'}^\mu & = & {x'}^\mu \left (x^\alpha \right ) .
\label{ap3}
\end{eqnarray}
\label{ap23}
\end{subequations}
The metric on the total space can be written as
\begin{subequations}
\begin{eqnarray}
ds^2_{(MD)} & = & 
b\left ( 
\omega^{\bar a} + h^{\bar a}_\mu dx^\mu
\right )
\left ( 
\omega_{\bar a} + h_{\bar a\mu} dx^\mu
\right ) + 
\left ( 
h^{\bar \mu}_\alpha dx^\alpha 
\right )
\left ( 
h_{\bar \mu\beta} dx^\beta 
\right )
\label{ap4}\\
\omega^{\bar a} & = & e^{\bar a}_b dy^b 
\quad
h^{\bar a}_b =  e^{\bar a}_b
\label{ap5}
\end{eqnarray}
\label{ap3-5}
\end{subequations}
where $x^\mu$ and $y^b$ are the coordinates along the base and fibres respectively; 
(Greek indices)$=0,1,2,3$ and (Latin indices)$=5,6,\cdots ,N$; 
$\bar A = \bar a,\bar \mu$ is the viel-bein index; 
$\eta_{\bar A\bar B} = \{\pm 1,\pm 1, \cdots ,\pm 1\}$
is the signature of the MD metric; $\omega^{\bar a}$ are 
the 1-forms satisfying to the structural equations 
\begin{equation}
d\omega^{\bar a} = f^{\bar a}_{\bar b\bar c}
\omega^{\bar b}\wedge\omega^{\bar c}
\label{ap6}
\end{equation}
where $f^{\bar a}_{\bar b\bar c}$ are the structural 
constants for the gauge group $\cal G$.
\par
The independent degrees of freedom for gravity on the 
principal bundle with the structural group ${\cal G}$ is 
vier-bein $h^{\bar\mu}_\nu(x^\alpha)$, gauge potential 
$h^{\bar a}_\nu(x^\alpha)$ and scalar field $b(x^\alpha)$ \cite{sal}. 
All functions depend only on the point $x^\mu$ on the base 
of the principal bundle as a consequence of the symmetry of the
fibres.
\par
Varying the action \eqref{ap1} with respect to $h^{\bar\mu}_\nu(x^\alpha)$ leads to 
\begin{equation}
\int\left (
R^\mu_{\bar\nu} - \frac{1}{2}h^\mu_{\bar\nu} R - 
\Lambda_1h^\mu_{\bar\nu} 
\right )
\sqrt {|\gamma|} d^Ny - \Lambda_2' h^\mu_{\bar\nu} = 0
\label{ap7}
\end{equation}
where $|\gamma| = \det h^{\bar a}_b = b^N\det e^{\bar a}_b$ 
is the volume element on the fibre and 
$\sqrt {|G|} = \sqrt {|g|}\sqrt {|\gamma|}$ is a consequence of 
the following structure of the MD metric
\begin{subequations}
\begin{eqnarray}
h & = & h^{\bar A}_B =  
\left (
\begin{array}{cc}
h^{\bar a}_b & h^{\bar a}_\mu \\
0 & h^{\bar \nu}_\mu
\end{array}
\right ) ,
\label{ap8} \\
h^{-1} & = & h^B_{\bar A} = 
\left (
\begin{array}{cc}
h^b_{\bar a} & -h^b_{\bar a}h^{\bar a}_\nu h^\nu_{\bar\nu} \\
0 & h^\mu_{\bar \nu}
\end{array}
\right ) ,  
\label{ap9} \\
h^b_{\bar a} & = & {\left (h^{\bar a}_b \right )} 
^{-1}
\quad
h^\mu_{\bar \nu} = {\left (h^{\bar \nu}_\mu \right )} 
^{-1} .
\label{ap10}
\end{eqnarray}
\label{ap10a}
\end{subequations}
An integration over the EDs can be easily performed since
no functions depend on $y^a$ 
\begin{equation}
\int \left (\cdots\right ) \sqrt {|\gamma|} d^Ny = 
\left (\cdots\right ) \int \sqrt {|\gamma|} d^Ny = 
\left (\cdots\right ) b^{N/2} V_{\cal G}
\label{ap11}
\end{equation}
where ${V_{\cal G}} = \int \sqrt{\det (e^{\bar a}_b)} d^Ny$  
is the volume of the gauge group $\cal G$. 
In this case Eq. \eqref{ap7} becomes
\begin{equation}
R^\mu_{\bar\nu} - \frac{1}{2}h^\mu_{\bar\nu} R = 
\left (
\Lambda_1 + \frac{\Lambda_2}{b^{N/2}}
\right )
h^\mu_{\bar\nu}
\label{ap12}
\end{equation}
where $\Lambda_2' = V_{\cal G}\Lambda_2$. 
\par
Varying with respect to $h^{\bar a}_\mu(x^\alpha)$ leads to 
\begin{equation}
R^\mu_{\bar a} = 0
\label{ap13}
\end{equation}
as $h^{\bar a}_\mu$ does not consists in 
$\det (h^{\bar A}_B) = \det (h^{\bar a}_b) \det (h^{\bar\mu}_\nu)$. 
\par
Varying with respect to $b(x^\alpha)$ leads to
\begin{equation}
\frac{\delta S}{\delta b} = \sum_{\bar a,b} 
\frac{\delta h^{\bar a}_b}{\delta b} \frac{\delta S}{\delta h^{\bar a}_b} 
= h^{\bar a}_A \left (
R^A_{\bar a} - \frac{1}{2} h^A_{\bar a} 
- \Lambda_1 h^A_{\bar a}
\right )
\label{ap14}
\end{equation}
here we used Eq. \eqref{ap12} and $h^\mu_{\bar a} = $. This 
equation we write in the form
\begin{equation}
R^{\bar a}_{\bar a} - \frac{N}{2} R = 
N \Lambda_1
\label{ap15}
\end{equation}
>From Eq. \eqref{ap12} we have 
\begin{subequations}
\begin{eqnarray}
h^{\bar\nu}_\mu
\left [
R^\mu_{\bar\nu} - \frac{1}{2} h^\mu_{\bar\nu} R - 
\left (
\Lambda_1 + \frac{\Lambda_2}{b^{N/2}}
\right )h^\mu_{\bar\nu} R
\right ] = 
h^{\bar\nu}_\mu \left [ \cdots \right ] + 
h^{\bar\nu}_a \left [ \cdots \right ] & = & 
\nonumber \\
h^{\bar\nu}_A
\left [
R^A_{\bar\nu} - \frac{1}{2} h^A_{\bar\nu} R - 
\left (
\Lambda_1 + \frac{\Lambda_2}{b^{N/2}}
\right )h^A_{\bar\nu} R
\right ] = 
R^{\bar\nu}_{\bar\nu} - 2 R - 4
 \left (
\Lambda_1 + \frac{\Lambda_2}{b^{N/2}}
\right ) & = & 0
\label{ap16}
\end{eqnarray}
\label{ap16a}
\end{subequations}
Adding Eqs. \eqref{ap16} and \eqref{ap15} we find
\begin{equation}
R = R^{\bar A}_{\bar A} = - \frac{2}{N+2}
\left [
\left ( N+4\right )\Lambda_1 + 
\frac{4\Lambda_2}{b^{N/2}}
\right ]
\label{ap17}
\end{equation}
Finally we have
\begin{subequations}
\begin{eqnarray}
R^{\bar a}_{\bar a} & = & -\frac{2N}{N+2} 
\left (
\Lambda_1 + \frac{\Lambda_2}{b^{N/2}}
\right ) ,
\label{ap18} \\
R^\mu_{\bar a} & = & 0
\label{ap19} \\
R^\mu_{\bar\nu} - \frac{1}{2}h^\mu_{\bar\nu} R & = & 
\left (
\Lambda_1 + \frac{\Lambda_2}{b^{N/2}}
\right )
h^\mu_{\bar\nu}
\label{ap20}
\end{eqnarray}
\label{ap20a}
\end{subequations}
This equation system can be rewritten as 
\begin{subequations}
\begin{eqnarray}
R^{\bar a}_{\bar a} & = & -\frac{2N}{N+2} 
\left (
\Lambda_1 + \frac{\Lambda_2}{b^{N/2}}
\right ) ,
\label{ap21} \\
R_{\bar\mu\bar a} & = & 0
\label{ap22} \\
R_{\bar\mu\bar\nu} - \frac{1}{2}\eta_{\bar\mu\bar\nu} R & = & 
\left (
\Lambda_1 + \frac{\Lambda_2}{b^{N/2}}
\right )
\eta_{\bar\mu\bar\nu}
\label{ap23a}
\end{eqnarray}
\label{ap23b}
\end{subequations}
here we have used $h^{\bar\nu}_b = 0$.

\end{document}